\documentclass{llncs}
\usepackage{multirow}
\usepackage{caption}
\usepackage{booktabs}
\usepackage{graphicx}
\usepackage{subcaption}
\usepackage{amsmath}
\usepackage{amssymb}
\usepackage{epsfig}
\usepackage{amsfonts}
\usepackage{longtable}
\usepackage{rotating}
\usepackage{color}
\usepackage{hyperref}

\hypersetup{
    colorlinks=true,
    linkcolor=blue,
    filecolor=magenta,
    urlcolor=blue,
}

\newcommand{\fig}{Fig.~}
\newcommand{\eq}{Eq.~}
\newcommand{\tab}{Table~}

\newcommand*{\tran}{^{\mkern-1.5mu\mathsf{T}}} 
\renewcommand{\vec}[1]{\underline{#1}}

\begin{document}
\frontmatter          
\pagenumbering{arabic}

\title{A distance-based loss for smooth and continuous skin layer segmentation in optoacoustic images.}

\author{Stefan Gerl\inst{1}\thanks{equal contribution} \and Johannes C. Paetzold\inst{2,4}$^{\star}$  \and Hailong He\inst{1,3,4}$^{\star}$  \and Ivan Ezhov\inst{2,4} \and Suprosanna Shit\inst{2,4} \and Florian Kofler\inst{2,4} \and Amirhossein Bayat \inst{2,4} \and Giles Tetteh \inst{2,4} \and Vasilis Ntziachristos \inst{1,3} \and Bjoern Menze\inst{2,4}}

\institute{Department of Electrical and Computer Engineering, Technische Universität München, Munich, Germany
\and Department of Computer Science, Technische Universität München, Munich, Germany
\and Institute of Biological and Medical Imaging (IBMI), Helmholtz Zentrum München, Neuherberg, Germany
\and TranslaTUM Center for Translational Cancer Research, Munich, Germany \\
\email{johannes.paetzold@tum.de}
}

\maketitle              

\begin{abstract} Raster-scan optoacoustic mesoscopy (RSOM) is a powerful, non-invasive optical imaging technique for functional, anatomical, and molecular skin and tissue analysis. 
However, both the manual and the automated analysis of such images are challenging, because the RSOM images have very low contrast, poor signal to noise ratio, and systematic overlaps between the absorption spectra of melanin and hemoglobin. Nonetheless, the segmentation of the epidermis layer is a crucial step for many downstream medical and diagnostic tasks, such as vessel segmentation or monitoring of cancer progression.
We propose a novel, shape-specific loss function that overcomes discontinuous segmentations and achieves smooth segmentation surfaces while preserving the same volumetric Dice and IoU. Further, we validate our epidermis segmentation through the sensitivity of vessel segmentation. We found a 20 $\%$ improvement in Dice for vessel segmentation tasks when the epidermis mask is provided as additional information to the vessel segmentation network. 
\end{abstract}

\section{Introduction}
Skin imaging plays an important role in dermatology; 
in both fundamental research and treatment of diverse diseases \cite{kittler2002diagnostic,zhang2018skin,patino2018automatic}. 
Optoacoustic (photoacoustic) mesoscopy offers unique 
opportunities in optical imaging, by bridging the 
gap between microscopic and macroscopic description of tissue and by enabling high-resolution visualizations which are deeper than optical microscopy \cite{rajpara2009systematic,anas2018towards}. Raster scan optoacoustic mesoscopy (RSOM) is a novel technique for noninvasive, high-resolution, and three-dimensional imaging of skin features based on optical absorption contrast \cite{omar2019optoacoustic,Aguirre2017}. Several studies using RSOM have recently demonstrated high resolution skin imaging by revealing different skin layers and the structure of the microvasculature \cite{Aguirre2017,Aguirre2014}. RSOM imaging has been used for in-depth visual examination of psoriasis and analysis of vascularization of superficial tumors \cite{Aguirre2017,omar2015pushing}. A critical first step for quantitative analysis of clinical RSOM images is to segment skin layers and vasculature in a rapid, reliable, and automated manner.\\ 
Previously, skin layers in RSOM images have been manually segmented by visual inspection of vasculature morphology; or automatically, based on signal intensity levels exploiting dynamic programming \cite{nitkunanantharajah2019skin} and random forest \cite{moustakidis2019fully}. Such procedures are slow, inaccurate, and unsuitable for processing larger numbers of patients, especially for making clinical decisions during the patient’s visit. Manual segmentation is also subjective and hence compromises the reproducibility and robustness of RSOM skin image analysis. In addition to the rich, three-dimensional vascular information, RSOM images can be employed to compute biomarkers such as the total blood volume, vessel density, and complexity. These help to assess disease progression and identify skin inflammation. In current practice, the segmentation of RSOM images is thresholding-based and thus very sensitive to signal to noise ratio (SNR) variations. 
Therefore, there is a need to develop a reliable, automatic skin layer and vessel segmentation approach based on neural networks for rapid quantitative analysis of RSOM images.
\begin{figure}
    \centering
    \begin{subfigure}{0.38 \textwidth}
        \centering  
        \includegraphics[height=4.2cm, trim={0 5cm 2cm 0},clip]{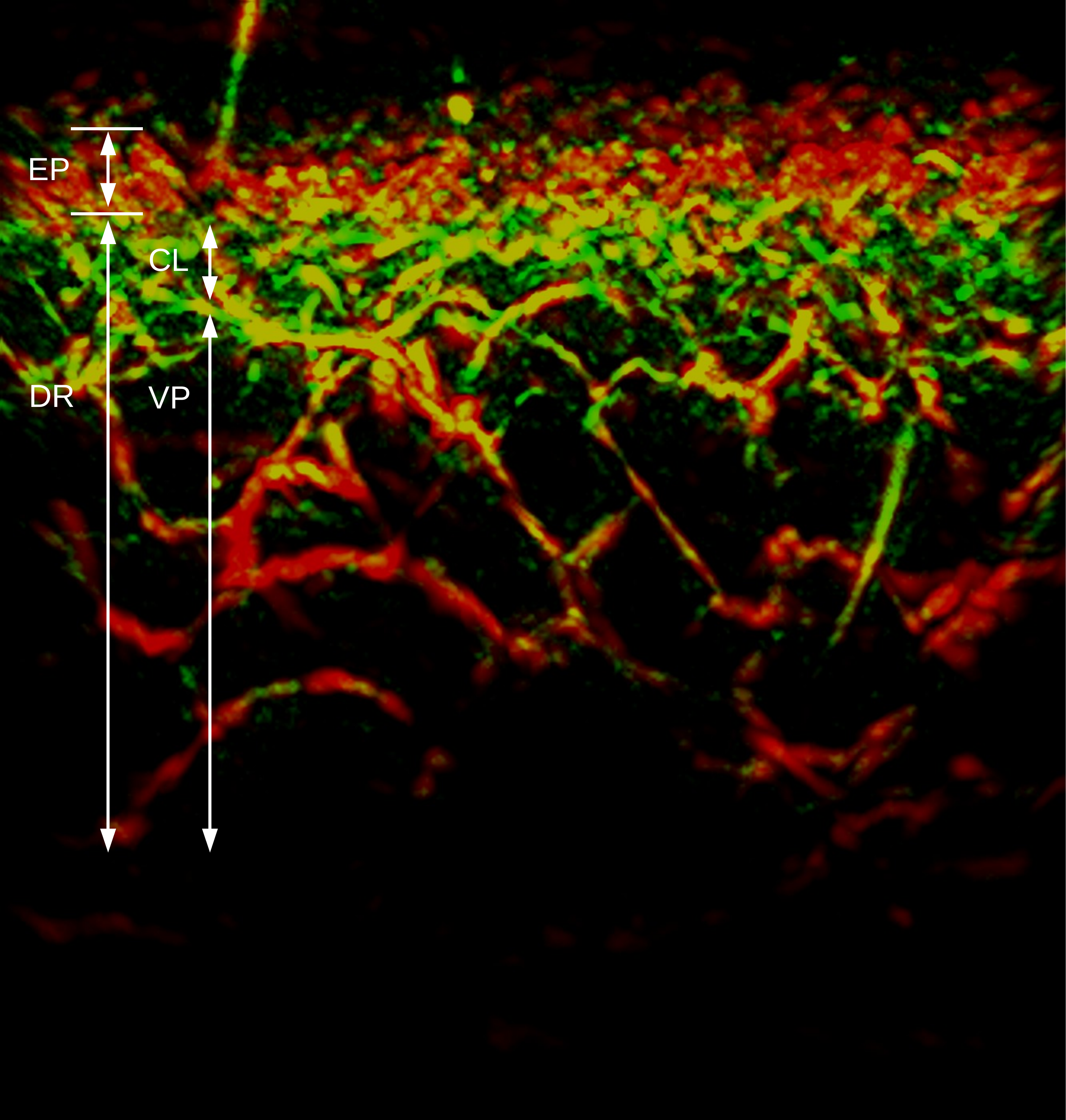}
        \caption{}        
    \end{subfigure}
    \begin{subfigure}{0.61\textwidth}
        \centering
        \includegraphics[height=4.2cm, trim={0 7cm 0 1.5cm},clip]{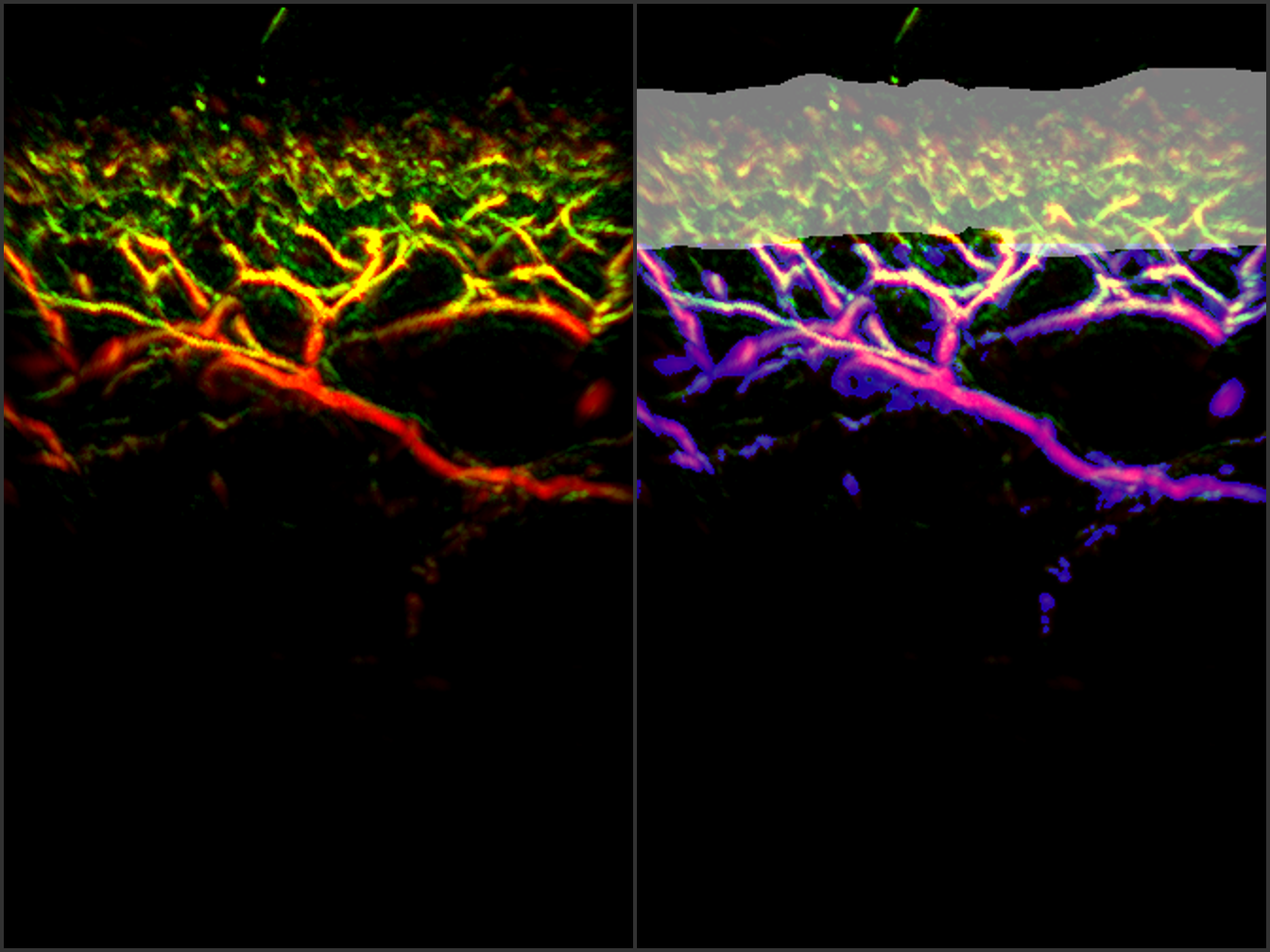}
        \caption{}
    \end{subfigure}
     \caption[Layers of RSOM visualized in MIP]
            {\textit{Visual problem definition:} (a) shows a maximum intensity projection (MIP) in $y$ direction of a volumetric RSOM 
            image of human skin. The anatomical structure is described by the white arrows: epidermis (EP) and dermis (DR); the dermis itself consists of the capillary loop layer (CL) on top of the vascular plexus (VP). Here, extracting an exact boundary is very difficult.
             (b) Our contribution: considering the RSOM image on the left of (b) we automated the epidermis segmentation (semitransparent white overlay on the right), which we use as a mask for the vascular segmentation of the vascular plexus (VP). The smoothness of the layer segmentation is crucial to input meaningful and reproducible images into the vessel segmentation. 
             }
    \label{fig:rsom_mip_example}
\end{figure}
\paragraph{\textbf{Problem Definition: }}
This paper sets out to develop a custom loss function for
structured and smooth epidermis segmentation in RSOM images, which can be used in any segmentation network. 
This is difficult for the RSOM modality because melanin and hemoglobin overlap in their acoustic response. I.e. it is hard to distinguish between the melanin layer from the epidermis and the capillary loops from the dermis layer, because the absorption spectrum of melanin and hemoglobin is very close at the used laser wavelength (532nm) of RSOM imaging \cite{Ntziachristos2010}. First, segmentation of the epidermis is necessary to compute the average thickness of the epidermis layer, which is an important biomarker. Second, a vessel segmentation is less affected by the melanin signal, if it is smoothly masked out, see Fig. \ref{fig:rsom_mip_example}. Critically, the response of melanin is distributed irregularly and nonlinear, which increases the difficulty of epidermis segmentation. The smoothness is the key aspect where traditional segmentation networks trained on, e.g., BCE or soft-dice loss fail, because their segmentations lead to discontinuous surfaces, which miss parts of the epidermis, see Fig. \ref{fig:dice_vs_s}. These "gaps" inevitably lead to false vessel segmentations because the melanin and hemoglobin signal cannot be distinguished, see Fig. \ref{fig:rsom_mip_example} and Supplementary Fig. 1.\\
Methodologically, we overcome this by developing a custom loss function; previous works demonstrated how custom loss functions can be superior for difficult medical imaging tasks \cite{al2017shape,hu2019topology,shit2020cldice,navarro2019shape}. 
Regarding smoothness, previous approaches used post-processing steps to achieve smooth surfaces, e.g., filters \cite{Manfredi2016SkinSR}.
More complex neural network approaches used topological concepts as priors for histology gland segmentation \cite{bentaieb2016topology}. For general smooth shape segmentation, other approaches \cite{kalogerakis20173d,ravishankar2017learning} successfully combined multiple fully convolutional networks, which incorporated arbitrary shape priors into the loss function of an additional network. Another successful approach used graph cuts \cite{srivastava2018three}. Patino et al. implemented superpixel merging \cite{patino2018automatic} and Li et al. graph theory to achieve smooth surfaces \cite{li2005optimal}.\\
\textit{\textbf{Our Contributions: }}At the core of our contribution is a new method to achieve an anatomically consistent and smooth epidermis layer segmentation in RSOM images. First, we introduce a custom loss term, which enforces smooth surfaces through a distance-based smoothness penalty. Next, we show that a combination of binary cross entropy loss and the custom smoothness loss optimizes epidermis segmentation. Conclusively, the resulting loss allows us to learn from very few examples, but well defined prior knowledge with very high accuracy, leading to the first automated RSOM epidermis segmentation algorithm, which preserves smooth layer structures. 
 We validate the epidermis layer segmentation by evaluating the performance of RSOM vessel segmentation - a downstream image processing task - with the proposed segmentation algorithm and its alternatives.

\section{Methodology}

\subsection{Loss function}

Our total loss function $\mathcal{L}_j:\mathbb{R}^{X\times Z}_{\ge 0} \rightarrow \mathbb{R}_{\ge 0}$ 
for a sample $j$ consists of a per-pixel cross-entropy part 
$\mathcal{H}_j:\mathbb{R}^{X\times Z}_{\ge 0}\rightarrow \mathbb{R}^{X\times Z}_{\ge 0}$ 
and a smoothness penalty
$\mathcal{S}_j:\mathbb{R}^{X\times Z}_{\ge 0}\rightarrow \mathbb{R}_{\ge 0}$,
which is weighted by a constant parameter
$s = \mathrm{const}$. The total loss function is given in Equation \ref{eq:total_loss}. 
The width and height of one 2D slice are $X$ and $Z$ (see also Section \ref{sec:dataset}),
consequently the summation term denotes the spatial average of $\mathcal{H}_j$.
\begin{equation}
    \mathcal{L}_j = \frac{1}{XZ} \sum_{x=1}^X \sum_{z=1}^Z \mathcal{H}_j\left(x,z\right) 
+ s\cdot \mathcal{S}_j
    \label{eq:total_loss}
\end{equation} 
$\mathcal{H}_j$ is a per-pixel standard binary cross entropy. 
To incorporate prior knowledge about the shape of the epidermis, the smoothness 
cost function $\mathcal{S}_j$ is defined in Equation \ref{eq:smoothness_loss}. 
This concept is motivated by the clinical imaging setup, where the epidermis layer is always approximately parallel to the $x-y$~plane. The scenario of arbitrary orientations, non-parallel to some coordinate plane would complicate implementation, but is implausible, as the RSOM scan is acquired directly and directional on the skin surface.
Firstly, we split the probability map $P_j \in \mathbb{R}^{X \times Z}_{\ge 0}$, in $Z$ row vectors:
\begin{equation} 
    P_j = 
    \begin{bmatrix}
        \vec{p}_{1,j}\tran,~ 
        \vec{p}_{2,j}\tran,~
        \hdots,~
        \vec{p}_{Z,j}\tran
    \end{bmatrix}.
    \label{eq:split_in_row_vec}
\end{equation}
Secondly, we perform a 1D convolution or correlation operation (denoted by 
$*$) of a vector $\vec{p}_{z,j}$ with kernel $K$, defined in Equation \ref{eq:conv_kernel}. 
Note that this is a discrete convolution and $K$ is a discrete kernel, 
and its weights are chosen to obey $\sum^{\infty}_{-\infty} K = 1$. Furthermore, 
convolving with $K$ does not change the size of $p_{z,j}$. 
In Equation \ref{eq:smoothness_loss}, $\oslash$ is the Hadamard division~\cite{Cyganek2013}, where $\mathbf{1}_{X}\in \mathbb{R}^X$ is a vector of ones. $\left| \cdot \right|$ denotes the element-wise absolute value. 
\begin{equation}
    \mathcal{S}_j = \sum_x \sum_z \left| \left(\left(\left( \vec{p}_{z,j}* K\right)+\mathbf{1} 
_{X}\right) \oslash \left(\vec{p}_{z,j} + \mathbf{1} _{X}\right) \right) -\mathbf{1} 
_{X} \right|
    \label{eq:smoothness_loss}
\end{equation}
\begin{equation}
    K\left(x\right) =
    \begin{cases}
        \frac{1}{5} & |x| \le 2 \\
        0 & \mathrm{else}
    \end{cases}
    \label{eq:conv_kernel}
\end{equation}

In the case of an equal prediction probability in $x$ direction, 
$p_{z,j} = c_z\cdot \mathbf{1}_{1,X}$, with $c_z \in \mathbb{R}_{\ge 0, \le 1}, c_z=\mathrm{const}$ 
for all $z$.
Consequently, $\mathcal{S}_j = 0$, which results in no smoothness penalty.\\
However, in the common case of an unequal prediction probability in $x$ direction, 
it follows that $\mathcal{S}_j > 0$; and $\mathcal{S}_j$ contributes
to the total loss function, i.e., penalizing a non-smooth layer in $x$ direction. 
Note that $\mathcal{S}_j$ is differentiable with respect to the model weights,
which is a necessary condition for any loss function. 
Due to incorporating the smoothness penalty, the model is directly taught 
to learn smooth representations, rather than requiring a manual post-processing step.
Note that the computation of $\mathcal{S}_j$ is very inexpensive, as it requires only one 1D convolution, additions, and one division.\\
In practice, a perfectly segmented healthy epidermis is not rectilinear across a whole RSOM image, as the thickness of the skin layers deviates in spatially coarse patterns. This means that the thickness differences are smooth and coarse but not abrupt, see Fig. \ref{fig:slice_loss_comparison}.
Thus, $\mathcal{S}_j > 0$, while at the same time $\sum_{x} \sum_z \mathcal{H}_j\left(x,z\right) 
= 0$, resulting in an overall nonzero loss.
Therefore, the scaling factor $s$ in Equation \ref{eq:total_loss} must be tuned accordingly.

\subsection{Network architectures}
We use two very general segmentation architectures to show that the novel loss function is agnostic to the network architecture. First, a U-Net\cite{Ronneberger2015} with dropout in all up-convolution blocks except the first one. Second, a fully convolutional network (FCN) with 7 layers depth and no dropout \cite{tetteh2018deepvesselnet}.  We train using the described 5 fold cross validation for all loss functions depicted in Table \ref{tab:scores1}. All networks are implemented in Pytorch using the Adam optimizer.

\section{Experiments and Discussion}
Since our objective is to achieve a smooth epidermis and dermis surface segmentation, while maintaining the accuracy of traditional overlap and volumetric scores (Dice, Precision, IoU), we compare the segmentation from a pure BCE loss function to our combined loss function for a starkly varying smoothness loss term, weighted by $s$.
We validate our epidermis segmentation by an additional sensitivity experiment. We run a standard CNN vessel segmentation on the masked image volume and show that the new and smooth epidermis segmentation is beneficial for vessel segmentation, thereby also for even further "downstream" tasks in clinical practice. 

\paragraph{\textbf{Dataset:}}
\label{sec:dataset}
The given RSOM dataset consists of two volumetric data channels with a size
of $333 \times 171 \times 500$ pixels
($X \times Y \times Z$). Step sizes are $\Delta x = \Delta y = 12~\mathrm{\mu m}$ and $\Delta z = 3~\mathrm{\mu m}$, resulting in image volumes of $2 \times 4 \times 1~\mathrm{mm^3}$, where part of the data can represent voxels outside the skin.
For the layer segmentation, data is processed in $x-z$ slices of $333 \times 500$ pixels.
We split our dataset consisting of 31 3D volumes according to these in 25 volumes for training and validation (5-fold cross validation) and 6 volumes for testing. 
Next, we split all 3D volumes along the $x-z$ slices and shuffle the train and validation set across the 25 volumes. Thereby, we have a training set of 3420 2D images ($20\times 171$), a validation set of 855 2D images ($5\times 171$), and a completely unseen test set of 1026 2D images ($6\times 171$). The GT of the epidermis was labeled using the approach in \cite{Aguirre2014,Aguirre2017} by experts familiar with RSOM images. In ambiguous situations, labels were discussed to reach consensus decisions.

\paragraph{\textbf{Assessment of the segmentation smoothness:}}
Dice score and IoU do not reveal detailed morphological information about the segmentation result.
In order to quantify the epidermis and dermis surface smoothness, we calculate
the arithmetic mean deviation in 1D, which is a common measure in material science to assess the quality of a surface \cite{Gok2012,Wiechec2016}. Where, $\tilde{\mu}_z\left(x\right)$, is the local mean at $x$ over a moving window of 5 pixels.\\
\begin{equation}
R_a = \frac{1}{X}\sum_{x=1}^X \left| z\left(x\right) - \tilde{\mu}_z\left(x\right) \right|
\end{equation}
The local mean respects the coarse structure of the skin layers reducing its contribution to the roughness $R_a$ to a minimum, all while the fine structure (high-frequency deviations) is reported. As an additional measure of the roughness, we use the angular distribution of surface normals, see Fig. \ref{fig:surface_normals}.
\newcommand{\txsc}{0.5}
\newcommand{\isc}{0.99}
\begin{figure}
    \centering
    \begin{subfigure}{\txsc\textwidth}
        \centering

        \includegraphics[width=\isc\textwidth]{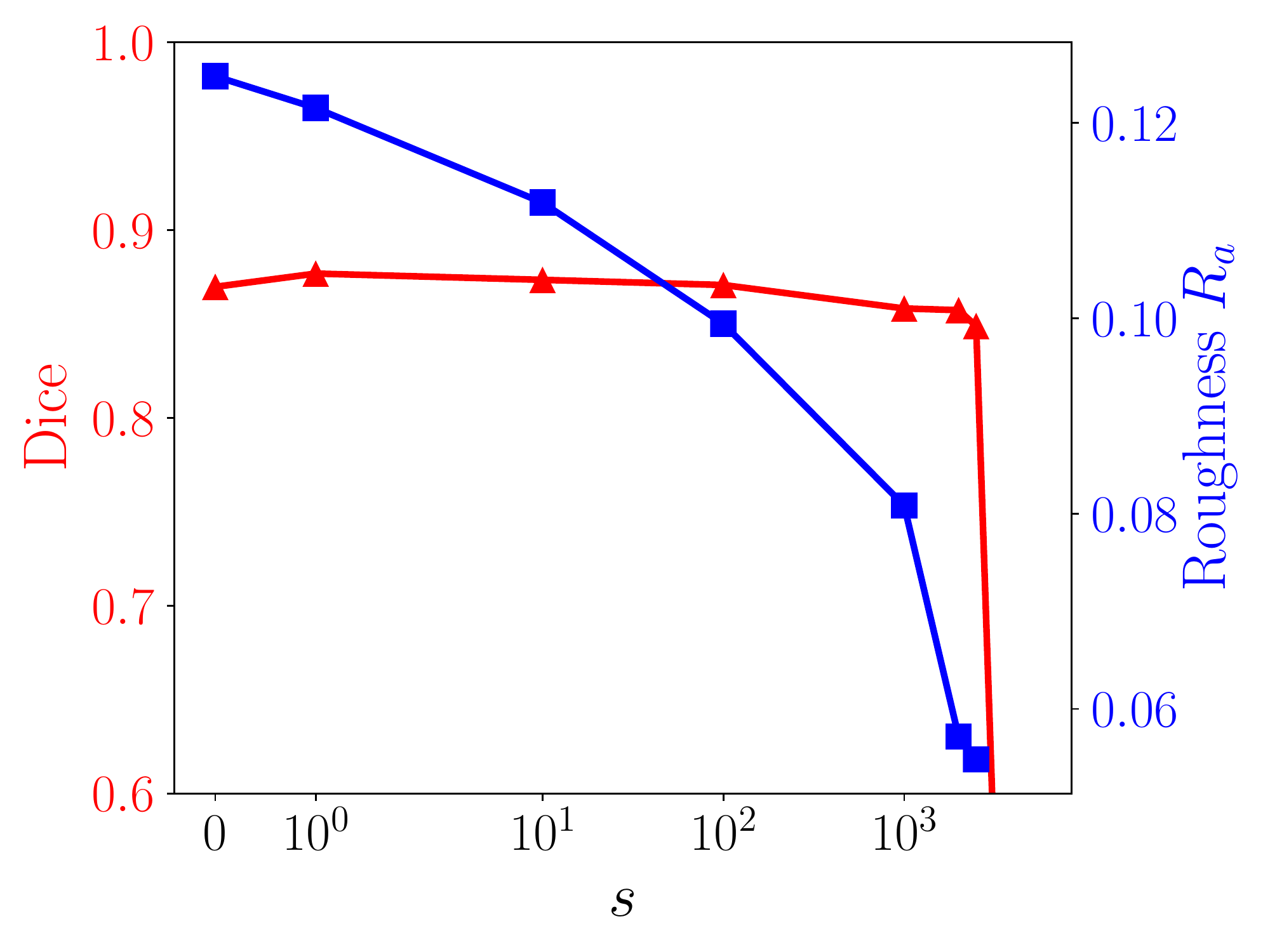}
        \caption{U-Net}
    \end{subfigure}%
    \begin{subfigure}{\txsc\textwidth}
        \centering
        \includegraphics[width=\isc\textwidth]{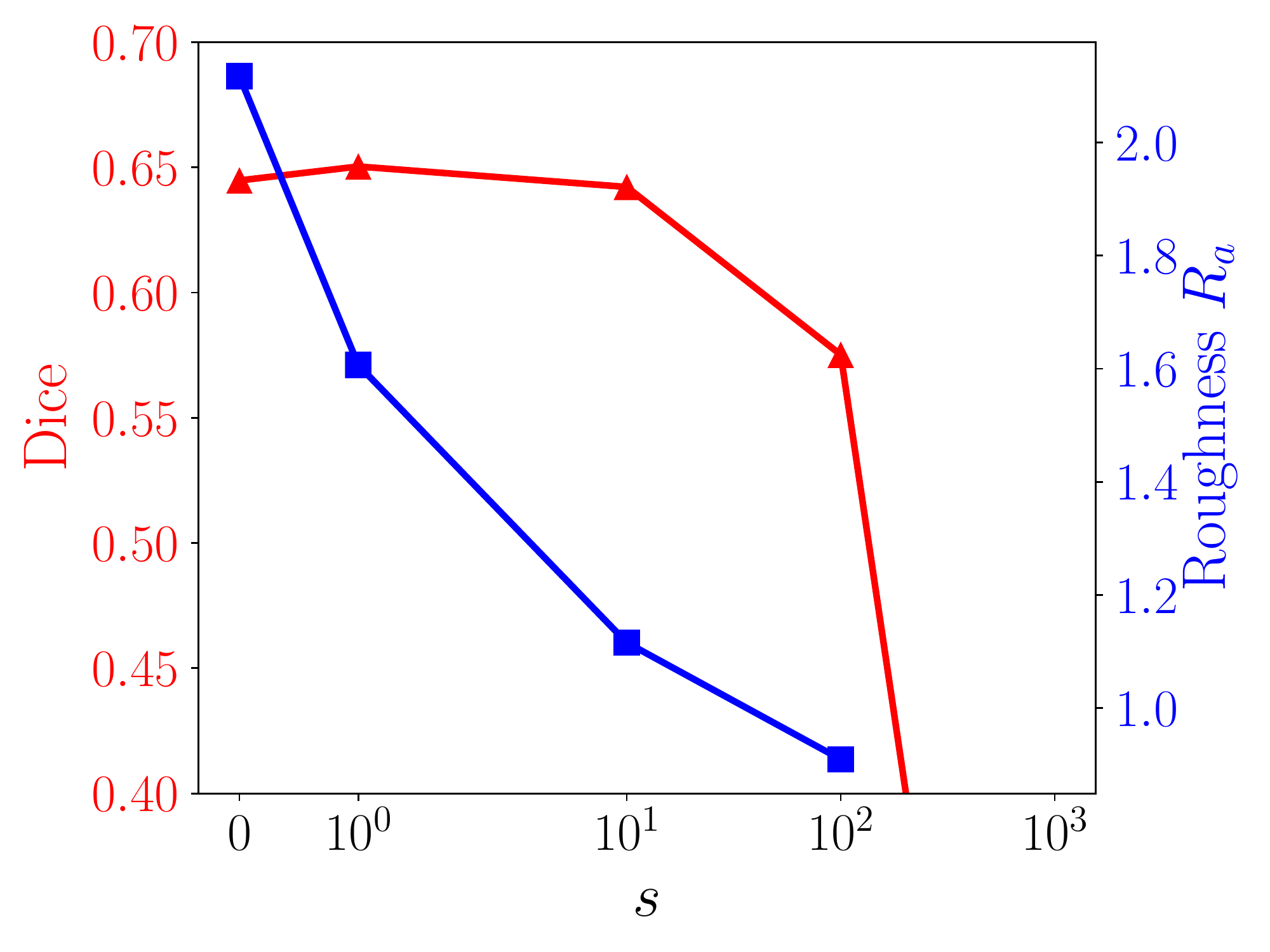}
        \caption{FCN}
    \end{subfigure}%
    \caption{The resulting Dice (red) and Roughness $R_a$ (blue) values for our Epidermis segmentation plotted against the scaling factor $s$ between
             BCE and smoothness loss in \eq\eqref{eq:smoothness_loss}. To calculate the roughness value, all 2D slices of the test set are accumulated; $R_a$ is calculated by averaging over all $x$, for all slices for both, the epidermis and the dermis surfaces. For both the U-Net and the FCN, increasing the smoothness loss substantially improves the surface roughness, while maintaining a robust Dice score, for a wide range of $s$ (1-2000). Please note that $s=0$ represents a pure BCE loss function. We consider this a strong property of our loss term, as its robustness across log-scales is evident.}
    \label{fig:dice_vs_s}
\end{figure}
\begin{table}
    \caption{\textit{Evaluation of the epidermis segmentation for the U-Net and FCN architecture for a varying $s$}. Overlap based scores, Dice, IoU, Precision, and Recall do not substantially differ for both U-Net and FCN when increasing the smoothness loss term $s$. In contrast, the surface roughness continuously improves with increasing $s$. Our U-Net outperforms the FCN in regards to overlap based scores and roughness as it is a substantially more complex model.}\label{tab:scores1}
    \centering
    \begin{tabular}{|c|c|c|c|c|c|c|c|}
        \hline
        Network & Loss & Smoothness & Dice & Precision & Recall & IoU & Roughn.  \\
                &       &factor $s$ &       &          &        &     & $R_a$           \\
        \hline
        U-Net   & BCE  & 0      & $0.87\pm 0.09$  & $0.87\pm 0.10$ & $0.89\pm 0.13$ & $0.78\pm 0.11$ & 0.125 \\
        U-Net   & BCE+S  & 1     & $0.88\pm 0.10$  & $0.88\pm 0.08$ & $0.90\pm 0.14$ & $0.79\pm 0.12$ & 0.122 \\
        U-Net   & BCE+S  & 10     & $0.87\pm 0.10$  & $0.87\pm 0.10$ & $0.89\pm 0.13$ & $0.79\pm 0.12$ & 0.112 \\
        U-Net   & BCE+S  & 100    & $0.87\pm 0.10$  & $0.86\pm 0.10$ & $0.90\pm 0.13$ & $0.78\pm 0.12$ & 0.099 \\
        U-Net   & BCE+S  & 1000   & $0.86\pm 0.09$  & $0.83\pm 0.11$ & $0.90\pm 0.12$ & $0.76\pm 0.11$ & 0.081 \\ 
        U-Net   & BCE+S  & 2000   & $0.86\pm 0.09$  & $0.80\pm 0.09$ & $0.94\pm 0.11$ & $0.76\pm 0.11$ & 0.057 \\
        U-Net   & BCE+S  & 2500   & $0.85\pm 0.11$  & $0.79\pm 0.11$ & $0.94\pm 0.12$ & $0.75\pm 0.13$ & 0.055 \\ 
        \hline
        7FCN    & BCE  & 0      & $0.64\pm 0.23$  & $0.85\pm 0.12$ & $0.62\pm 0.32$ & $0.52\pm 0.24$ & 2.127 \\
        7FCN    & BCE+S  & 1      & $0.65\pm 0.23$  & $0.85\pm 0.12$ & $0.63\pm 0.32$ & $0.52\pm 0.23$ & 1.607 \\
        7FCN    & BCE+S  & 10     & $0.64\pm 0.24$  & $0.85\pm 0.12$ & $0.62\pm 0.32$ & $0.52\pm 0.24$ & 1.116 \\
        7FCN    & BCE+S  & 100    & $0.58\pm 0.27$  & $0.83\pm 0.14$ & $0.55\pm 0.35$ & $0.45\pm 0.26$ & 0.909 \\
        \hline
    \end{tabular}
\end{table}
\newcommand{\txscb}{0.33}
\newcommand{\iscb}{0.80}
\begin{figure}
    \centering
    \begin{subfigure}{\txscb\textwidth}
        \centering
        
        \includegraphics[width=\iscb\textwidth]{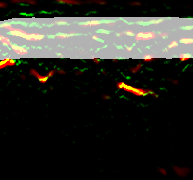}
        \caption{}
    \end{subfigure}%
    \begin{subfigure}{\txscb\textwidth}
        \centering
        \includegraphics[width=\iscb\textwidth]{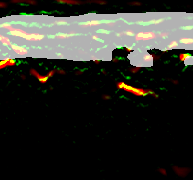}
        \caption{}
    \end{subfigure}%
    \begin{subfigure}{\txscb\textwidth}
        \centering
        \includegraphics[width=\iscb\textwidth]{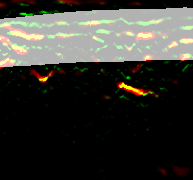}
        \caption{}
    \end{subfigure}
    \caption{\textit{Magnified slice of an RSOM skin scan}. The epidermis layer
             is marked in white. (a) ground truth annotation (label), (b) segmentation result
             from the U-Net with BCE loss, (c) segmentation result from the U-Net with
             BCE and Smoothness Loss ($s=2000$). Note that despite the highly unevenly distributed
             melanin response, the smoothness loss based prediction segments the epidermis layer superior with a very smooth surface, which is similar to the label .}
    \label{fig:slice_loss_comparison}
\end{figure}
\\ \textit{\textbf{Epidermis segmentation: }}We train the U-Net and FCN architectures incorporating loss functions with differing smoothness terms. Inclusion of our novel loss term in any proportion improves the smoothness of the layer segmentation, as measured by $R_a$, independent of the network architecture, see Table \ref{tab:scores1}. The same trend is visible in the distribution of surface normals, see Fig. \ref{fig:surface_normals}. Dice scores are insensitive to the magnitude of the smoothness term, defined by $s$, until certain tipping points, where the networks fail to converge, see Fig. \ref{fig:dice_vs_s}. Two-sided, paired Wilcoxon signed rank tests, comparing Dice scores for different smoothness factors to the pure BCE loss, support this observation by revealing no significant difference in underlying distributions for both U-Nets and FCNs with p-values $>$0.05 across the board (p-values $>$ 0.4 for all U-Net models and $>$ 0.06 for all FCN models). On the other hand, p-values for $R_a$ show that our models have significantly different $R_a$ distributions across test samples (p-values $<$ 0.05 for all FCN and all U-Net where s $>$ 1). 
Overall, we achieve the best performance of around 87 \% Dice and 0.06 $R_a$ using the U-Net architecture.
Across the samples (for the U-Net), the scores of the five-fold cross validation resulted in an agreement of the Dice scores of $0.985\pm0.00068$. From the very low standard deviation, we conclude that the statistical divergence between the training and validation set is very low. Visual inspection reveals that using our smooth loss indeed yields smooth and continuous epidermis and dermis surfaces, see Fig. \ref{fig:rsom_mip_example} (b), Fig. \ref{fig:slice_loss_comparison} and Supplementary Fig. 3.
The combined loss is robust for a largely varying $s$. For the U-Net, the loss was stable for $s$ ranging from 1 to 2800 and for the FCN for an $s$ ranging from 1 to 1000. To be clear, our smoothness term is not a standalone loss, but works very well in combination with BCE; increasing the factor $s$ too much leads to instabilities during training, see \fig\ref{fig:dice_vs_s}. \\
\newcommand{\txsca}{0.333}
\newcommand{\isca}{0.9}
\begin{figure}[t]
    \centering
    \begin{subfigure}{\txsca\textwidth}
        \centering
        \includegraphics[width=\isca\textwidth]{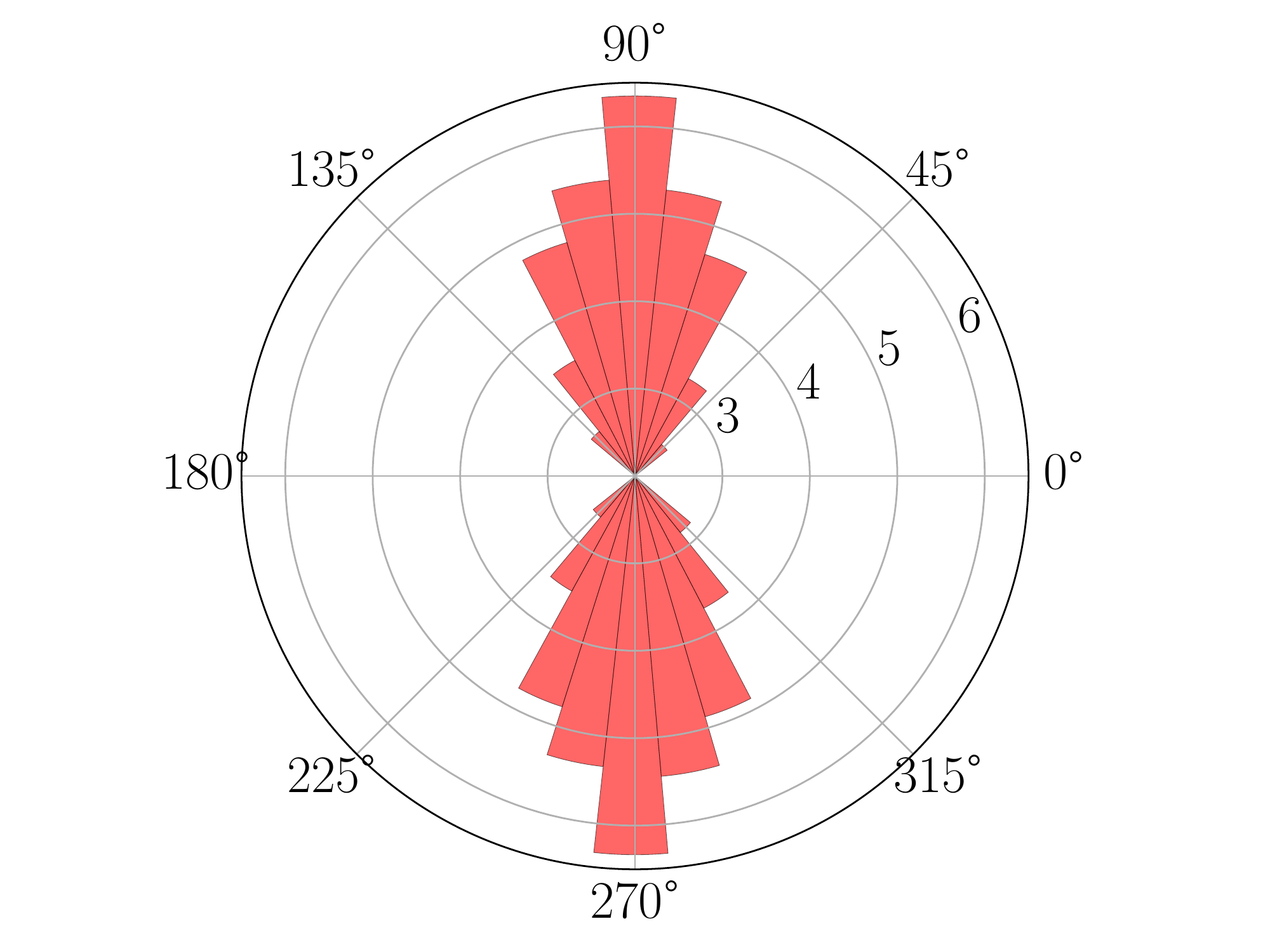}
        \caption{}
    \end{subfigure}%
    \begin{subfigure}{\txsca\textwidth}
        \centering
        \includegraphics[width=\isca\textwidth]{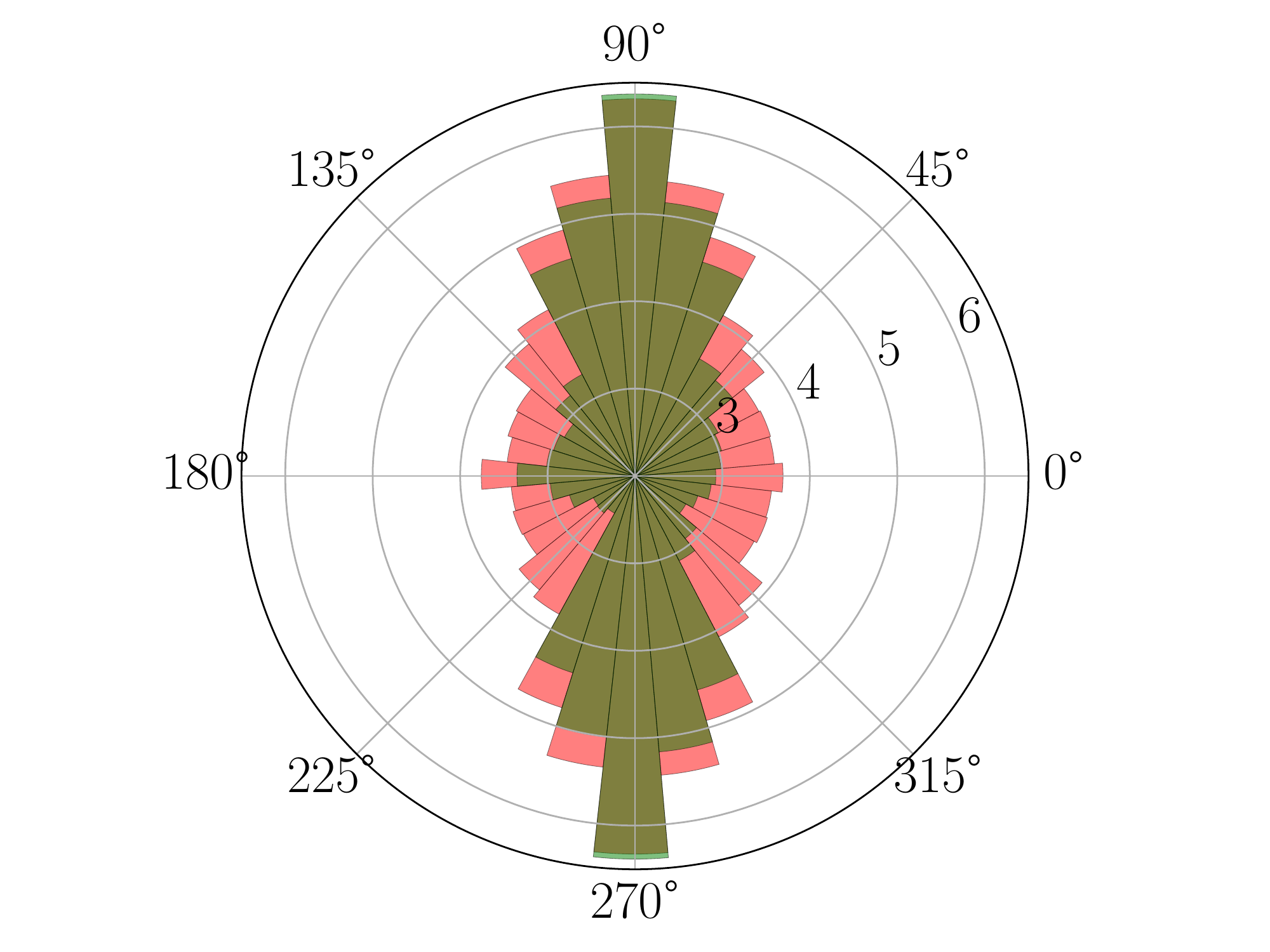}
        \caption{}
    \end{subfigure}%
    \begin{subfigure}{\txsca\textwidth}
        \centering
        \includegraphics[width=\isca\textwidth]{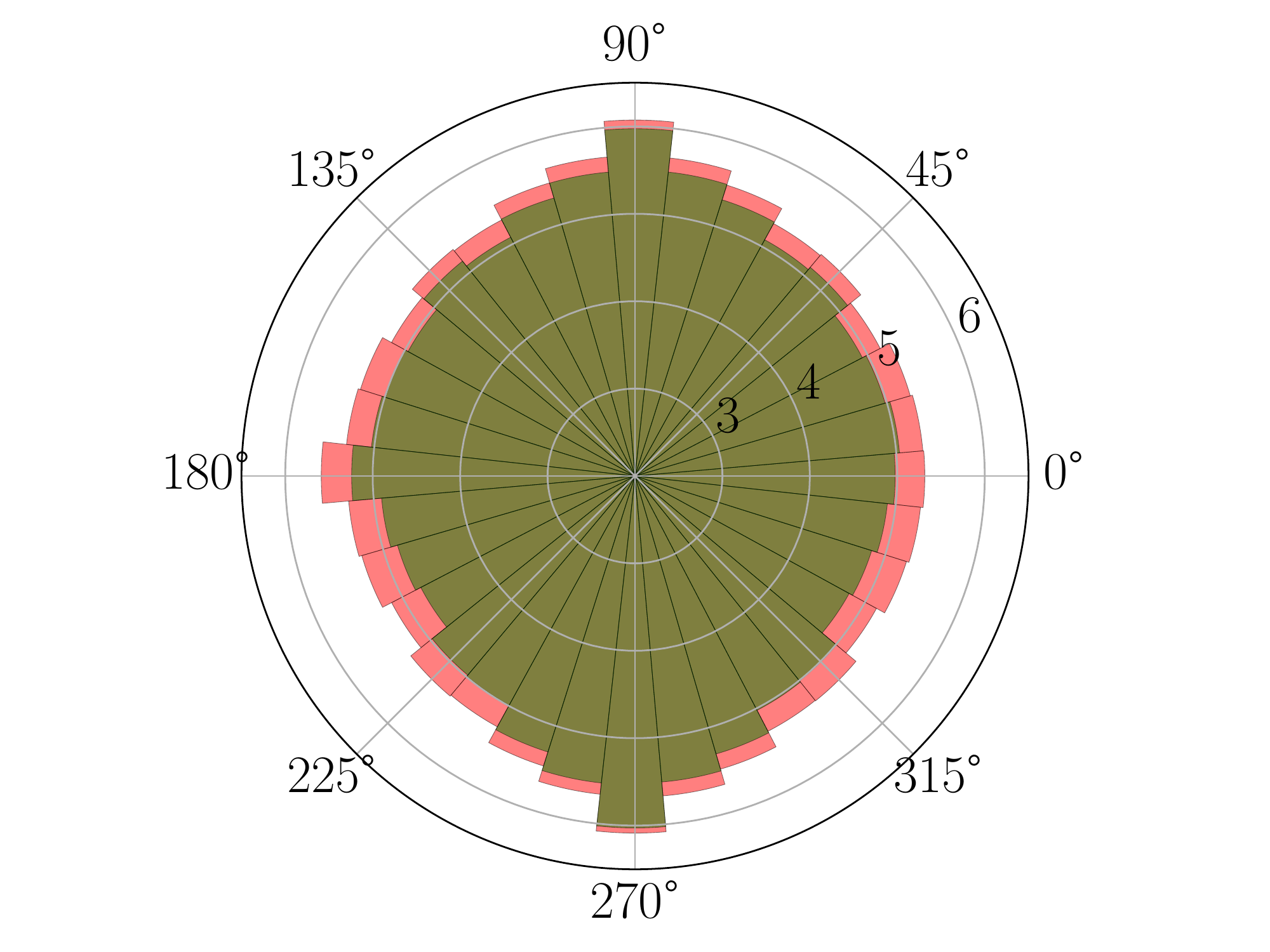}
        \caption{}
    \end{subfigure}
    \caption{\textit{Histogram of the orientation of surface normals on the segmentation map}
             (logarithmic scale). (a) Ground truth. 
             BCE loss (red), BCE and smoothness loss (green) for (b) U-Net ($s = 2000$), and 
             (c) FCN ($s = 100$). The amount of wrongly orientated surface normals is reduced by one order of magnitude for the U-Net and less pronounced for the FCN, too. The anatomically desired smooth surface is better segmented using our smoothness loss term. For details on the calculation of the surface normals, please see the supplementary material.}
    \label{fig:surface_normals}
\end{figure}
\begin{table}
\centering
\caption{\textit{Vessel segmentation sensitivity experiment: }Here we report the performance of a standard vessel segmentation using DeepVesselNet \cite{tetteh2018deepvesselnet}; without using our proposed epidermis layer segmentation and using our method. Numbers in bold indicate superior performance.}
\label{tab:scores2}
\begin{tabular}{|c|c|c|c|c|} 
\hline
Configuration & Dice                     & Precision                & Recall                   & IoU                       \\ 
\hline
no Mask       & 0.619$\pm$0.187          & 0.673$\pm$0.303          & 0.698$\pm$0.160          & 0.474$\pm$0.196           \\ 
\hline
our Method    & \textbf{0.810$\pm$0.095} & \textbf{0.883$\pm$0.100} & \textbf{0.760$\pm$0.125} & \textbf{0.690$\pm$0.117}  \\
\hline
\end{tabular}
\end{table}

\paragraph{\textbf{Vessel Segmentation:}}
We validate our epidermis segmentation via a sensitivity experiment of vessel segmentation, where we use the epidermis segmentation as a mask. Vessel segmentation in optoacoustic skin scans is of great clinical interest in order to
characterize the vasculature of healthy human skin and in order to diagnose several disease cases, where vasculature and
capillaries are altered or damaged; e.g., for the diagnosis of long term effects of diabetes on the patients' body. An established method \cite{tetteh2018deepvesselnet,todorov2019automated,paetzold2019transfer} for vessel segmentation is used to verify the validity of the epidermis segmentation. Synthetically generated arterial trees \cite{schneider2012tissue,schneider2015joint} serve as training and validation data, see Supplementary Fig. 2. Testing is done on $32$ annotated 3D RSOM volumes of size $166 \times 85 \times 250$. Epidermis segmentation increased the Dice similarity by more than 20 \% from $0.619\pm0.187$ to $0.810\pm0.095$, yielding high confidence of the validity and necessity of our epidermis segmentation approach. The complete results for vessel segmentation are given in~\tab\ref{tab:scores2}.

\section{Conclusion}
In this paper, we introduced a novel, shape-specific loss function, for RSOM image skin layer segmentation. Our loss overcomes discontinuous segmentations and achieves smooth segmentation surfaces, while preserving the same volumetric segmentation performance, e.g. Dice. This is important because only meaningful and reproducible segmentation can be used for downstream tasks in medical practice, e.g. vessel segmentation for diagnostic purposes. 
We validate our epidermis segmentation through a sensitivity experiment, where we use our epidermis segmentation as a mask for vessel segmentation and improve their performance by more than 20 \% Dice. 

{\bibliographystyle{splncs03}
\bibliography{egbib}

\begin{thebibliography}{10}
\providecommand{\url}[1]{\texttt{#1}}
\providecommand{\urlprefix}{URL }

\bibitem{Aguirre2017}
Aguirre, J., Schwarz, M., Garzorz, N., Omar, M., Buehler, A., Eyerich, K.,
  Ntziachristos, V.: Precision assessment of label-free psoriasis biomarkers
  with ultra-broadband optoacoustic mesoscopy. Nature Biomedical Engineering
  1(5),  0068 (2017), \url{https://doi.org/10.1038/s41551-017-0068}

\bibitem{Aguirre2014}
Aguirre, J., Schwarz, M., Soliman, D., Buehler, A., Omar, M., Ntziachristos,
  V.: Broadband mesoscopic optoacoustic tomography reveals skin layers. Opt.
  Lett.  39(21),  6297--6300 (Nov 2014),
  \url{http://ol.osa.org/abstract.cfm?URI=ol-39-21-6297}

\bibitem{al2017shape}
Al~Arif, S.M.R., Knapp, K., Slabaugh, G.: Shape-aware deep convolutional neural
  network for vertebrae segmentation. In: International Workshop and Challenge
  on Computational Methods and Clinical Applications in Musculoskeletal
  Imaging. pp. 12--24. Springer (2017)

\bibitem{anas2018towards}
Anas, E.M.A., Zhang, H.K., Kang, J., Boctor, E.M.: Towards a fast and safe
  led-based photoacoustic imaging using deep convolutional neural network. In:
  International Conference on Medical Image Computing and Computer-Assisted
  Intervention. pp. 159--167. Springer (2018)

\bibitem{bentaieb2016topology}
BenTaieb, A., Hamarneh, G.: Topology aware fully convolutional networks for
  histology gland segmentation. In: International Conference on Medical Image
  Computing and Computer-Assisted Intervention. pp. 460--468. Springer (2016)

\bibitem{Cyganek2013}
Cyganek, B.: Tensor Methods in Computer Vision, chap.~2, pp. 9--188. John Wiley
  \& Sons, Ltd (2013)

\bibitem{Gok2012}
Gok, A., Gologlu, C., Demirci, H., Kurt, M.: Determination of surface qualities
  on inclined surface machining with acoustic sound pressure. Strojniski
  Vestnik  58,  587--597 (06 2012)

\bibitem{hu2019topology}
Hu, X., Li, F., Samaras, D., et~al.: Topology-preserving deep image
  segmentation. In: Advances in Neural Information Processing Systems. pp.
  5658--5669 (2019)

\bibitem{kalogerakis20173d}
Kalogerakis, E., Averkiou, M., Maji, S., Chaudhuri, S.: 3d shape segmentation
  with projective convolutional networks. In: Proceedings of the IEEE
  Conference on Computer Vision and Pattern Recognition. pp. 3779--3788 (2017)

\bibitem{kittler2002diagnostic}
Kittler, H., Pehamberger, H., Wolff, K., Binder, M.: Diagnostic accuracy of
  dermoscopy. The lancet oncology  3(3),  159--165 (2002)

\bibitem{li2005optimal}
Li, K., Wu, X., Chen, D.Z., Sonka, M.: Optimal surface segmentation in
  volumetric images-a graph-theoretic approach. IEEE transactions on pattern
  analysis and machine intelligence  28(1),  119--134 (2005)

\bibitem{Manfredi2016SkinSR}
Manfredi, M., Grana, C., et~al.: Skin surface reconstruction and 3d vessels
  segmentation in speckle variance optical coherence tomography. In: VISIGRAPP
  (2016)

\bibitem{moustakidis2019fully}
Moustakidis, S., Omar, M., Aguirre, J., Mohajerani, P., Ntziachristos, V.:
  Fully automated identification of skin morphology in raster-scan optoacoustic
  mesoscopy using artificial intelligence. Medical physics  46(9),  4046--4056
  (2019)

\bibitem{navarro2019shape}
Navarro, F., Shit, S., Ezhov, I., Paetzold, J., Gafita, A., et~al.: Shape-aware
  complementary-task learning for multi-organ segmentation. In: International
  Workshop on Machine Learning in Medical Imaging. pp. 620--627. Springer
  (2019)

\bibitem{nitkunanantharajah2019skin}
Nitkunanantharajah, S., Zahnd, G., Olivo, M., Navab, N., Mohajerani, P.,
  Ntziachristos, V.: Skin surface detection in 3d optoacoustic mesoscopy based
  on dynamic programming. IEEE transactions on medical imaging  (2019)

\bibitem{Ntziachristos2010}
Ntziachristos, V., Razansky, D.: Molecular imaging by means of multispectral
  optoacoustic tomography (msot). Chemical Reviews  110(5),  2783--2794 (2010),
  \url{https://doi.org/10.1021/cr9002566}, pMID: 20387910

\bibitem{omar2019optoacoustic}
Omar, M., Aguirre, J., Ntziachristos, V.: Optoacoustic mesoscopy for
  biomedicine. Nature biomedical engineering  3(5),  354--370 (2019)

\bibitem{omar2015pushing}
Omar, M., Schwarz, M., Soliman, D., Symvoulidis, P., Ntziachristos, V.: Pushing
  the optical imaging limits of cancer with multi-frequency-band raster-scan
  optoacoustic mesoscopy (rsom). Neoplasia  17(2),  208--214 (2015)

\bibitem{paetzold2019transfer}
Paetzold, J.C., Schoppe, O., Al-Maskari, R., Tetteh, G., Efremov, V., Todorov,
  M.I., Cai, R., et~al.: Transfer learning from synthetic data reduces need for
  labels to segment brain vasculature and neural pathways in 3d. In:
  International Conference on Medical Imaging with Deep Learning--Extended
  Abstract Track (2019)

\bibitem{patino2018automatic}
Pati{\~n}o, D., Avenda{\~n}o, J., Branch, J.W.: Automatic skin lesion
  segmentation on dermoscopic images by the means of superpixel merging. In:
  International Conference on Medical Image Computing and Computer-Assisted
  Intervention. pp. 728--736. Springer (2018)

\bibitem{rajpara2009systematic}
Rajpara, S., Botello, A., Townend, J., Ormerod, A.: Systematic review of
  dermoscopy and digital dermoscopy/artificial intelligence for the diagnosis
  of melanoma. British Journal of Dermatology  161(3),  591--604 (2009)

\bibitem{ravishankar2017learning}
Ravishankar, H., Venkataramani, R., et~al.: Learning and incorporating shape
  models for semantic segmentation. In: International conference on medical
  image computing and computer-assisted intervention. pp. 203--211. Springer
  (2017)

\bibitem{Ronneberger2015}
Ronneberger, O., Fischer, P., Brox, T.: U-net: Convolutional networks for
  biomedical image segmentation (2015)

\bibitem{schneider2015joint}
Schneider, M., Hirsch, S., Weber, B., Sz{\'e}kely, G., Menze, B.H.: Joint 3-d
  vessel segmentation and centerline extraction using oblique hough forests
  with steerable filters. Medical image analysis  19(1),  220--249 (2015)

\bibitem{schneider2012tissue}
Schneider, M., Reichold, J., Weber, B., Sz{\'e}kely, G., Hirsch, S.: Tissue
  metabolism driven arterial tree generation. Medical image analysis  16(7),
  1397--1414 (2012)

\bibitem{shit2020cldice}
Shit, S., Paetzold, J.C., Sekuboyina, A., Zhylka, A., Ezhov, I., Unger, A.,
  Pluim, J.P., Tetteh, G., Menze, B.H.: cldice--a topology-preserving loss
  function for tubular structure segmentation. arXiv preprint arXiv:2003.07311
  (2020)

\bibitem{srivastava2018three}
Srivastava, R., Yow, A.P., Cheng, J., Wong, D.W., Tey, H.L.: Three-dimensional
  graph-based skin layer segmentation in optical coherence tomography images
  for roughness estimation. Biomedical Optics Express  9(8),  3590--3606 (2018)

\bibitem{tetteh2018deepvesselnet}
Tetteh, G., Efremov, V., Forkert, N.D., Schneider, M., Kirschke, J., et~al.:
  Deepvesselnet: Vessel segmentation, centerline prediction, and bifurcation
  detection in 3-d angiographic volumes. arXiv preprint arXiv:1803.09340
  (2018)

\bibitem{todorov2019automated}
Todorov, M.I., Paetzold, J.C., Schoppe, O., Tetteh, G., Shit, S., Efremov, V.,
  Todorov-V{\"o}lgyi, K., D{\"u}ring, M., Dichgans, M., Piraud, M., et~al.:
  Machine learning analysis of whole mouse brain vasculature. Nature Methods
  17(4),  442--449 (2020)

\bibitem{Wiechec2016}
Wiecheć, A., Nowicka, K., Błażewicz, M., Kwiatek, W.: Effect of magnetite
  composite on the amount of double strand breaks induced with x-rays. Acta
  Physica Polonica A  129,  174--175 (02 2016)

\bibitem{zhang2018skin}
Zhang, J., Xie, Y., Wu, Q., Xia, Y.: Skin lesion classification in dermoscopy
  images using synergic deep learning. In: International Conference on Medical
  Image Computing and Computer-Assisted Intervention. pp. 12--20. Springer
  (2018)

\end{thebibliography}
}
\end{document}